\input epsf.sty
\documentclass[preprint,amsmath,amssymb]{revtex4}
\usepackage{graphicx}
\begin{document}

\title{The bang of a white hole in the early universe from a 6D vacuum state:
Origin of astrophysical spectrum}
\author{Mauricio Bellini\footnote{E-mail address: mbellini@mdp.edu.ar}}

\affiliation{Departamento de F\'{\i}sica, Facultad de Ciencias
Exactas y Naturales, Universidad Nacional de Mar del Plata, Funes
3350, (7600) Mar del Plata,
Argentina.\\
Consejo Nacional de Investigaciones Cient\'{\i}ficas y T\'ecnicas
(CONICET), Argentina}

\begin{abstract}
Using a previously introduced model in which the expansion of the
universe is driven by a single scalar field subject to
gravitational attraction induced by a white hole during the
expansion (from a 6D vacuum state), we study the origin of squared
inflaton fluctuations spectrum on astrophysical scales.
\end{abstract}

\maketitle
\section{Introduction}

In a previous work\cite{ultimo}, we 
introduced a new formalism where, instead of implementing
a dynamical foliation by taking a spatial dependence of the fifth 
coordinate including its time dependence, we considered another
extra dimension, the sixth dimension, making possible the
implementation of two dynamical foliations in a sequential manner.
The first one was considered by choosing the fifth coordinate
depending of the cosmic time, and the second one by choosing the
sixth coordinate as dependent of the 3D spatial coordinates (in
our case considered as isotropic). Of course, all of these choices
preserve the continuity of the metric. In addition, the 6D metric
must be Ricci-flat. This requirement is a natural extension of the
vacuum condition used in the STM theory\cite{PR}, in which 5D
Ricci-flat metrics are used\cite{wesson} and the cylinder
condition has been eliminated in favor of retaining the metric's
dependence on the extra coordinate. In simple words, we used the
Campbell-Magaard theorem\cite{kachiporra} and its extensions for
embedding a 5D Ricci-flat space-time in a 6D Ricci-flat
space-time.
The conditions of 6D 
Ricci-flatness and the continuity of the metric gives us the
foliation of the sixth coordinate. These conditions specify the
sixth dimension as a function of the 3D spatial coordinates, in
order to establish the foliation. This function, for a particular
6D metric, can be seen in 4D as a gravitational potential related
to a localized
compact object 
that has the characteristics of a white hole. From a more general
point of view,
this is a mechanism 
for inducing localized matter onto a time-varying 4D hypersurface
by establishing a spatial foliation of a sixth coordinate from a
6D Ricci-flat metric. The importance of this approach lies in
that it can describe matter at both, 
cosmological and astrophysical scales, in an expanding universe.
The use of 6D physics is currently popular in particle
physics\cite{6d}, but with
the sixth dimension as time-like, rather than space-like.\\

In this letter we aim to study some predictions of this model on
astrophysical scales. The power spectrum of matter is one of the
most important statistics to describe the large-scale and
astrophysical-scale structures of the universe. The studies
developed in the last years have shown that on astrophysical
scales the power spectrum of galaxies and clusters of galaxies can
be satisfactorily expressed by a power law with an index between
$-1.9$ and $-1.5$\cite{einasto}. On larger scales the spectrum
turns over reaching a maximum on scales of $(100 - 150) \ h^{-1} \
Mpc$. We shall suppose that the evolution of structure in the
universe is only due to gravity.

\section{Review of the formalism}

\subsection{Effective 4D dynamics from a 6D vacuum state}

In order to describe a 6D vacuum, we consider the recently
introduced 6D Riemann-flat metric\cite{ultimo}
\begin{equation}\label{ini}
dS^2 = \psi^2 dN^2 - \psi^2 e^{2N} \left[dr^2+ r^2 d\Omega^2\right]
-d\psi^2 - d\sigma^2
\end{equation}
which defines a 6D vacuum state $G_{ab}=0$ ($a,b=0,1,2,3,4,5$). We
consider the 3D spatial space in spherical coordinates: $\vec r
\equiv \vec r(r,\theta,\phi)$; here $d\Omega^2 = d\theta^2 + {\rm
sin}^2(\theta) d\phi^2$. The metric (\ref{ini}) resembles the 5D
Ponce de Leon one\cite{pdl}, but with one additional space-like
dimension. Furthermore, the coordinate $N$ is dimensionless and
the extra (space-like) coordinates $\psi$ and $\sigma$ are
considered as noncompact. We define a physical vacuum state on the
metric (\ref{ini}) through the action for a scalar field
$\varphi$, which is nonminimally coupled to gravity
\begin{equation}\label{ac}
I = {\Large\int} d^6x \left[ \frac{^{(6)} {\cal R}}{16\pi G}
+ \frac{1}{2} g^{ab} \varphi_{,a} \varphi_{,b} -
\frac{\xi}{2} {^{(6)} {\cal R}} \varphi^2\right],
\end{equation}
where $^{(6)} {\cal R}=0$ is the Ricci scalar
and $\xi $ gives the coupling of $\varphi$ with gravity.
Implementing the coordinate transformation $N=H t$ and $R=r\psi = r/H$ on the frame
$U^{\psi} = (d\psi / dS)=0$ (considering $H$ as a
constant), followed by the foliation $\psi = H^{-1}$ on the metric (\ref{ini}), we obtain the effective 5D metric
\begin{equation}\label{5d}
^{(5)} dS^2 = dt^2 - e^{2 H t} \left[dR^2 + R^2 d\Omega^2\right] -
d\sigma^2,
\end{equation}
which is not Ricci-flat because $^{(5)} {\cal R}= 12 H^2$.
However, it becomes Riemann-flat in the limit $H 
\rightarrow 0$ i.e. $\left.R^A_{BCD}\right|_{H\rightarrow 0} =0$,
so that in this limit a 5D vacuum given by $\left. G_{AB}
\right|_{H\rightarrow 0} =0$, ($A, B =0,1,2,3,4$). Hence, we can
take the foliation $d\sigma^2 = 2 \Phi_n(R) \  dR^2$ in the metric
(\ref{5d}) on the sixth coordinate, and we obtain the effective 4D
metric
\begin{equation}\label{4d}
^{(4)} dS^2 = dt^2 - e^{2 H t} \left[ (1+ 2\Phi_n(R)) dR^2 + R^2
d\Omega^2\right],
\end{equation}
where $t$ is the cosmic time, and $H=\dot a/a$ is the Hubble
parameter for the scale factor $a(t) = a_0 e^{H t}$, with $a_0 =
a(t=0)$. The Einstein equations for the effective 4D metric
(\ref{4d}) are $G_{\mu\nu} = - 8 \pi G \  T_{\mu\nu}$
($\mu,\nu=0,1,2,3$), where $T_{\mu\nu}$ is represented by a
perfect fluid: $T_{\mu\nu} = ({\rm p} + \rho) u_{\mu} u_{\nu} -
g_{\mu\nu} {\rm p}$, ${\rm p}$ and $\rho$ being the pressure and
the energy density on the effective 4D metric (\ref{4d}). In a
previous work\cite{ultimo} $\Phi_n(R)$ was found for a puntual
mass $M_n=n M_p/3$ ($M_p=1.2 \times 10^{19} \  {\rm GeV}$ is the
Planckian mass) located at $R=0$, in the absence of expansion ($H
\rightarrow 0$)
\begin{equation}\label{sphi}
\Phi_n(R) = \frac{-3 G M_n {\rm ln}(R/R_*)}{R+6 G M_n {\rm
ln}(R/R_*)}.
\end{equation}
Here, $R_*$ is the value of $R$ such that $\Phi_n(R_*)=0$ and
$G=M^{-2}_p$ the gravitational constant. Hence, the function
$\Phi_n(R)$ describes the geometrical deformation of the metric
induced from a 5D flat metric, by a mass $M_n$. This function is
$\Phi_n > 0$ (or $\Phi_n < 0$) for $R < R_*$ ($R > R_*$),
respectively. Furthermore, $\left.\Phi_n(R)\right|_{R\rightarrow
\infty} \rightarrow 0$, and thereby the effective 4D metric
(\ref{4d}) is (in their 3D ordinary spatial components)
asymptotically flat. In this analysis we are considering the usual
4-velocities $u^{\alpha} = (1,0,0,0)$. The equation of state for a
given $R$ is\cite{ultimo}
\begin{equation}\label{state}
\frac{{\rm p}}{\rho} = -1 -\frac{2 G M_n}{R^3} \frac{\left[ 1-{\rm
ln}(R/R_*)\right] e^{-2 H t}}{ \left[H^2 - \frac{2 G M_n}{R^3}
e^{-2 H t}\right] },
\end{equation}
being ${\rm p} = {\rm p}_R + {\rm p}_{\theta} + {\rm p}_{\phi}$.
From the equation (\ref{state}) we can see that at the end of
inflation, when the number of e-folds is sufficiently large, the
second term in (\ref{state}) becomes negligible on cosmological
scales [on the infrared (IR) sector], and the equation of state on
this sector describes an asymptotic vacuum dominated (inflationary
) expansion:
\begin{equation}\label{state1} \left.{\rm
p}\right|^{(end)}_{IR} \simeq - \left.\rho\right|^{(end)}_{IR}.
\end{equation}

On the other hand, for $t=0$, we obtain (on arbitrary scales)
\begin{equation}
\left.\frac{{\rm p}}{\rho}\right|_{t=0} = - \frac{\left[H^2 -
\frac{2G M_n}{R^3} {\rm ln}\left( R/R_*\right)\right]}{\left[ H^2
- \frac{2GM_n}{R^3}\right]},
\end{equation}
which for a non-expanding universe ($H \rightarrow 0$) gives us,
for $R < R_*$, the equation of state for primordial galaxies
\begin{equation}
\left.\frac{{\rm p}}{\rho}\right|_{H\rightarrow 0,t=0} \simeq {\rm
ln}\left(R_*/R\right),
\end{equation}
which means that primordial galaxy formation should be possible on
scales smaller than $R_*$.

\begin{figure*}
\includegraphics[totalheight=8.5cm]{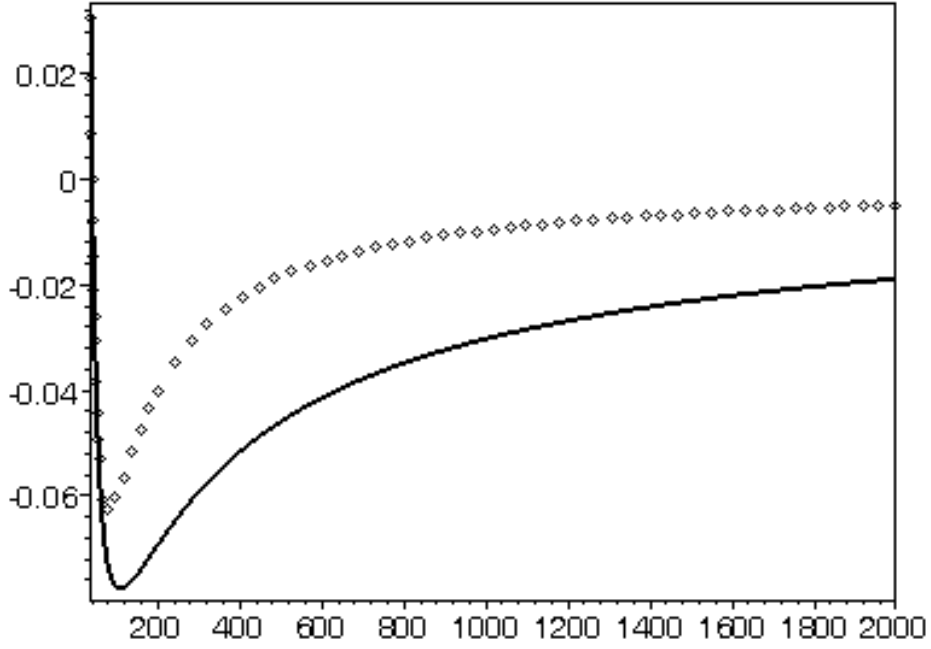}
\caption{\label{f1} We show the functions $\Phi_n(R)$ as a
function of $0.9 \  R_* < R < 50 \  R_*$ ($R_* = 4 n \lambda_p$
--- $\lambda_p$ is the Planckian wavelength) in their exact (continuous line)
and approximated versions
(pointed line), corresponding to the eqs. (\ref{sphi}) and
(\ref{6}), respectively. We use the values $n=10$, $l=0$ and $G=1$
(only in the figure).}
\end{figure*}

The effective 4D action for the universe is
\begin{equation}\label{4}
^{(4)} I = {\Large\int} d^4x \left[\frac{^{(4)} {\cal R}}{16 \pi
G} + \frac{1}{2} g^{\mu\nu} \varphi_{,\mu} \varphi_{,\nu} -
\frac{\xi_1}{2} \left.^{(4)} {\cal R}\right|_{\Phi_n=0} \varphi^2
- \frac{\xi^{n,l}_2(R)}{2} \left.^{(4)} {\cal R}\right|_{H=0}
e^{-2 H t} \varphi^2 \right],
\end{equation}
where $^{(4)} {\cal R} = 12 H^2 - (12 G M_n/ R^3) e^{-2 H t}$ is
the effective 4D Ricci scalar for the effective 4D metric
(\ref{4d}), $\xi_1$ gives the coupling of the scalar field
$\varphi(\vec R,t)$ with gravity on the background induced by the
foliation of the first extra dimension $\psi$ at $\Phi_n(R)=0$ and
$\xi^{n,l}_2(R)$ gives the coupling of $\varphi$ with gravity, on
the background induced by the foliation of the second extra
dimension $\sigma$ at $H=0$. The equation of motion for the field
$\varphi$ on the metric (\ref{4d}) is
\begin{eqnarray}
&&\ddot\varphi + 3 H \dot\varphi + e^{-2 H t} \left[
\frac{1}{R^2(1+2\Phi_n(R))} \frac{\partial}{\partial R} \left[ R^2
\frac{\partial\varphi}{\partial R} \right] - \frac{1}{(1+
2\Phi_n(R))^2} \frac{\partial\Phi_n}{\partial R}
\frac{\partial \varphi}{\partial R} \right. \nonumber \\
&& \left.+ \frac{1}{R^2 {\rm sin}(\theta)} \frac{\partial}{\partial\theta}
\left[{\rm sin}(\theta) \frac{\partial\varphi}{\partial\theta}\right]
\right. \nonumber \\
&&+\left. \frac{1}{R^2 {\rm sin}^2(\theta)}
\frac{\partial^2\varphi}{\partial\phi^2} \right] + \left[ \xi_1
\left.^{(4)}{\cal R}\right|_{\Phi_n=0} + \xi^{n,l}_2(R)
\left.^{(4)} {\cal R}\right|_{H=0} e^{-2 H t}\right] \varphi
=0.\label{5}
\end{eqnarray}
Notice that under this approach, the expansion is affected by a
geometrical deformation induced by the gravitational attraction of
a white hole of mass $M_n =n M_p/3$. This deformation is described
by the function $\Phi_n(R)$, which tends to zero on cosmological
scales [see figure (\ref{f1})].

\subsection{Weak field approximation}

In this section we study the weak field approximation for the
equation (\ref{5}). In that limit approximation the function
$\Phi_n(R)$ in (\ref{sphi}) can be written as
\begin{equation}\label{6}
\Phi_n(R) \simeq - \frac{3 GM_n}{R} + \left(\frac{6 G
M_n}{R}\right)^2,
\end{equation}
$M_n$ being the mass of the compact object located at $R=0$. Note
that the function (\ref{6}), as well as the exact one
(\ref{sphi}), goes to zero at $R\rightarrow \infty$. For $M_n>0$
there is a stable equilibrium for test
particles at $R_{*}=12 G M_n$ and it exhibits a gravitational 
repulsion (antigravity) for $R< R_{*}$. Hence, this object has the
properties of a white hole \cite{KL}. In the figure (\ref{f1}) we
have plotted $\Phi_n(R)$ given by the eqs. (\ref{sphi}) [the exact
expression plotted with a continuous line] and (\ref{6}) [the weak
field approximated expression plotted with a pointed line],
respectively, for $R > R_*$. Notice that the difference between
both is more important on smaller scales. Furthermore, it is
evident that the exact expression of $\Phi_n(R)$ is more sensitive
to the interaction. In order to obtain solutions of the equation
(\ref{5}) we propose $\varphi(\vec R,t) \sim \varphi_{t}(t)
\varphi_R(R) \varphi_{\theta,\phi}(\theta, \phi)$. With this
choice and using the fact that [see the eq. (11) in
\cite{ultimo})]: $-3G M_n = {R^2 d\Phi_n/dR \over (1+2\Phi_n)^2} +
{R\Phi_n \over 1+2\Phi_n}$, we obtain
\begin{eqnarray}
\ddot\varphi_t + 3 H \varphi_t - 12 H^2 \xi_1 \varphi_t &=&
-\alpha_t \varphi_t
e^{-2 H t}, \label{1a} \\
\frac{\partial}{\partial R} \left[ R^2 \frac{\partial\varphi_R}{
\partial R} \right] &+& \left[ 3 G M_n (1+2\Phi_n) + R\Phi_n\right]
\frac{\partial \varphi_R}{\partial R} \nonumber \\
&=& \varphi_R \left[ \frac{12 G M_n}{R} \xi^{n,l}_2(R) -\alpha_t
R^2 + \alpha_R\right]
(1+2\Phi_n), \label{1b} \\
{\rm sin}(\theta) \frac{\partial}{\partial\theta} \left[ {\rm
sin}(\theta) \frac{\partial \varphi_{\theta,\phi}}{
\partial\theta}\right] &+& \frac{1}{{\rm sin}^2(\theta)} \frac{
\partial^2\varphi_{\theta,\phi}}{\partial\phi^2} = \alpha_R
\varphi_{\theta,\phi}, \label{1c}
\end{eqnarray}
where $\alpha_t$ and $\alpha_R$ are separation constants. The
solution of the equation (\ref{1c}) is given by the spherical
harmonics $Y_{l,m}(\theta,\phi)$ [$\alpha_R=l(l+1)$]
\begin{equation}
\varphi_{\theta,\phi}(\theta,\phi)\sim \sum_{l,m} A_{lm} \
 Y_{l,m}(\theta,\phi) = \sum_{l,m} A_{lm} \
\sqrt{\frac{(2l+1)(l-m)!}{4\pi (l+m)!}} \  {\cal P}^m_l({\rm
cos}(\theta),
\end{equation}
where $m=-l,-(l-1),...,0,...,(l-1),l$ is a separation constant and
${\cal P}^m_l({\rm cos}(\theta)$ are the Legendre polynomials:
$P^m_l(x) = [(-1)^m / (2^l l!)] (1-x^2)^{m/2} (d^{l+m}/ dx^{l+m})
(x^2-1)^l$.

In order to study astrophysical implications for the solutions of
eqs. (\ref{1a}), (\ref{1b}) and (\ref{1c}), we shall concentrate
on the dispersive case ($\alpha_t = |\alpha_t|=k^2_R$) for
$\varphi(\vec R,t)$. In this case $k_R$ is the wavenumber related
to the coordinate $R$. On the other hand, for solving the equation
(\ref{1b}), we propose
\begin{equation}
\varphi_R(R) = \varphi_{k_R}(R) \  e^{-\int f_n(R) dR},
\end{equation}
such that $\varphi_{k_R}(R) = \varphi_R[\Phi_n=0]$. With this
choice (\ref{1b}) can be replaced by the equations
\begin{eqnarray}
&& \frac{d^2\varphi_{k_R}}{dR^2} + \frac{2}{R} \frac{d\varphi_{k_R}}{dR} +
\varphi_{k_R} \left[k^2_R - \frac{l(l+1)}{R^2}\right]=0, \label{3a} \\
&& \frac{d\varphi_{k_R}}{dR} \left[6 G M_n(1+2\Phi_n) + 2R\Phi_n\right]
+\varphi_{k_R} \left[ R^2 \left(f^2_n - \frac{df_n}{dR}\right) - 2 R f_n
+ 2\Phi_n \left(k^2_R R^2 - l(l+1)\right) \right.\nonumber \\
&& - \left. 3 G M_n (1+2\Phi_n) f_n - R \Phi_n f_n - \frac{12 G
M_n \xi^{n,l}_2(R)}{R} \left(1+2\Phi_n \right)\right]=0.
\label{3b}
\end{eqnarray}
The solution for the equation (\ref{3a}) is
\begin{equation}\label{3c}
\varphi_{k_R}(R) =
\frac{A}{\sqrt{R}} \  {\cal J}_{l+1/2}[k_R R],
\end{equation}
where ${\cal J}_{l+1/2}$ are the Bessel functions.
On the other hand, from the equation (\ref{3b}) we obtain the coupling
$\xi^{n,l}_2(R)$
\begin{eqnarray}
\xi^{n,l}_2(R) & = & \frac{R}{12 G M_n (1+2\Phi_n)} \left\{
\frac{1}{\varphi_{k_R}}\frac{d\varphi_{k_R}}{dR} \left[ 6 G M_n
(1+2\Phi_n) + 2 R \Phi_n\right] - \left[ R^2\left(f^2_n -
\frac{df_n}{dR}\right)
\right. \right. \nonumber \\
&-& \left.\left. 2 R f_n + 2\Phi_n \left[k^2_R R^2 - l(l+1)\right]
-3 G M_n \left( 1+2\Phi_n \right) f_n - R \Phi_n
f_n\right]\right\},   \label{coup}
\end{eqnarray}
with $f_n(R) = -[1/ (2R^2)] \left[3 G M_n \left(1+ 2\Phi_n(R)\right)
+ R \Phi_n(R)\right]$.
The solution of the equation (\ref{1a}) for the dispersive case is
\begin{equation}                  \label{26}
\varphi_t(t) \sim \xi_{k_R}(t)
= A \  e^{-3 H t/2} {\cal H}^{(2)}_{\nu}\left[\frac{k_R}{H} e^{-H t}\right],
\end{equation}
$A$ being a constant, $\nu = (1/2)\sqrt{9 + 48 \xi_1}$ and ${\cal
H}^{(2)}_{\nu}$ is the Hankel function of second kind. The
complete solution for the field $\varphi(\vec R,t)$ can be written
as
\begin{eqnarray}
\varphi(\vec R, t) &=& \frac{1}{(2\pi)^{3/2}}
e^{-\left[\left(\frac{n\lambda_p}{2R}\right)^2
\left(1+\frac{16 n \lambda_p}{3R}\right)\right]}
{\Large\int} d^3 k_R \sum^{l}_{m=-l} \sum^{n-1}_{l=0} \left[
a_{k_R l m} Y_{l,m}(\theta,\phi) \varphi_{k_R}(R)
\xi_{k_R}(t) \right. \nonumber \\
&+& \left. a^{\dagger}_{k_R l m} Y^*_{l,m}(\theta,\phi)
\varphi^*_{k_R}(R) \xi^*_{k_R}(t)\right], \label{var}
\end{eqnarray}
where $\varphi_{k_R}(R)$ and $\xi_{k_R}(t)$ are given respectively
by the expressions (\ref{3c}) and (\ref{26}).

\section{Astrophysical-scale spectrum}

In this section we shall study the structure formation during
inflation on astrophysical scales. In the framework of
cosmological scales, we understand that these are small scales, on
which one feels the presence of the compact object [and thus we
shall consider $\Phi_n(R) \neq 0$]. We must understand the present
astrophysical scales ($\sim$ 100 Mpc). We shall refer to this part
of the spectrum as the small-scale (SS) spectrum. It is known from
observation that galaxies and clusters of galaxies are correlated.
This should be responsible for differences in the clustering
properties of the populations in the nearby universe. One
population is characteristic for rich superclusters, and the other
for poorer ones. The former population has a power spectrum with a
sharp peak and a correlation function with zero crossing near $60
\  h^{-1} \ {\rm Mpc}$. The later population has a flatter power
spectrum and a zero crossing of the correlation function near $40
\  h^{-1} \ {\rm Mpc}$.

The  effective small-scale squared fluctuations in presence of
$\Phi_n(R)\neq 0$ are given by[see \cite{ultimo}]
\begin{equation}\label{sma3}
\left<\varphi^{2}\right>_{SS}^{(\Phi_n(R)\neq 0)}=\frac{1}{2\pi^2}
e^{-\left[\frac{1}{2}\left(\frac{n\lambda
_{p}}{R}\right)^2\left(1+\frac{16n\lambda _{p}
}{3 R}\right)\right]} \int _{\epsilon k_{H}}^{k_{H}}\frac{dk_{R}}{k_{R}}k_{R}^{3}\left.\left[\xi _{k_R}\xi 
_{k_R}^{*}\right]\right|_{IR}^{(\Phi_n(R)\simeq 0)}
\end{equation}
where $\epsilon =(k_{max}^{(IR)}/k_p)\ll 1$ is a dimensionless
constant parameter, and
$k_{max}^{IR}=\left.k_{H}(t_{i})=\sqrt{12\xi _{1}+(9/4)}H e^{Ht}
\right|_{t=t_i}$ is the wavenumber related with the Hubble radius
at the time $t_i$ (when the horizon re-enters),  and $k_{p}$ is
the Planckian wavenumber. The asymptotic expression for the modes
on the IR sector is\cite{ultimo}
\begin{equation}\label{modosIR}
\left.\xi_{k_R}(t) \right|_{IR}^{(\Phi_n(R)\simeq 0)} \simeq
-\frac{i}{2} \sqrt{\frac{1}{\pi H}} \Gamma(\nu)
\left(\frac{k_{\bar R}}{2H}\right)^{-\nu} e^{H t (\nu-3/2)}.
\end{equation}
Inserting (\ref{modosIR}) into (\ref{sma3}) we obtain
\begin{equation}\label{sma4}
\left<\varphi^{2}\right>_{SS}^{(\Phi_n(R)\neq
0)}=\frac{2^{2\nu-3}}{3-2\nu}\frac{\Gamma^{2}(\nu)}{\pi^3}
H^{2\nu-1}e^{-(3-2\nu)Ht}
e^{-\left[\frac{1}{2}\left(\frac{n\lambda
_{p}}{R}\right)^2\left(1+\frac{16n\lambda _{p} }{3
R}\right)\right]} k_{H}^{3-2\nu}(1-\epsilon)^{3-2\nu}.
\end{equation}

\begin{figure*}
\includegraphics[totalheight=8.5cm]{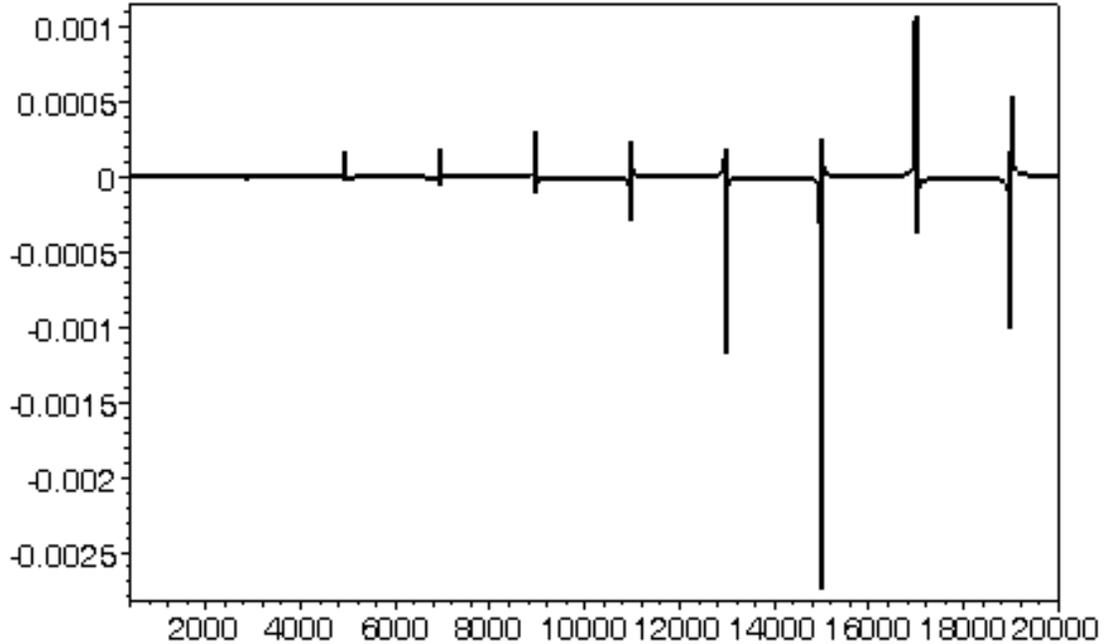}
\caption{\label{f2} The figure shows the coupling $\xi^{n,l}_2(R)$
as a function of $10 \ R_* <R< 500 \ R_*$ ($R_* = 4 n \lambda_p$
--- $\lambda_p$ is the Planckian wavelength). We use the values
$n=10$, $l=0$ and $G=1$ (only in the figure).}
\end{figure*}

\begin{figure*}
\includegraphics[totalheight=8.5cm]{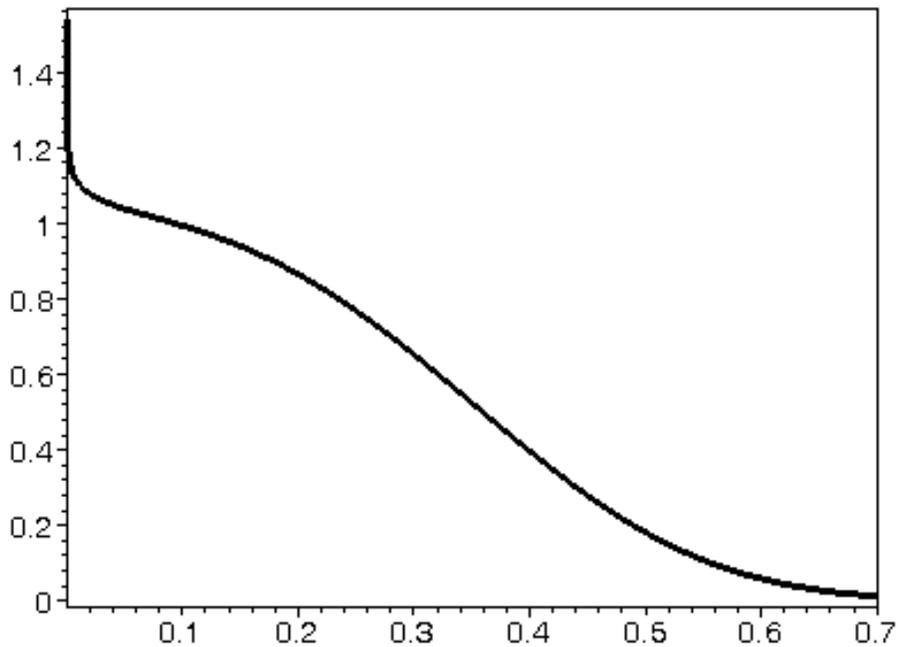}
\caption{\label{f3} The figure shows the power spectrum ${\cal
P}(k_R)$ of $\left<\varphi^2\right>_{SS}$ as a function of $0.0001
< k_R < 0.7$ (we take the Planckian wavenumber value as $k_p = 2
\pi $). We use the values $n=10$, $l=0$, $G=1$ (only in the
figure) and $n_s=0.964$ (which conrresponds to $\nu=1.518$).}
\end{figure*}

Hence, the expression  (\ref{sma4}) becomes
\begin{equation}
\left<\varphi^{2}\right>_{SS}^{(\Phi_n(R)\neq 0)} \simeq
e^{-\left[\frac{1}{2}\left(\frac{n\lambda
_{p}}{R}\right)^2\left(1+\frac{16n\lambda _{p} }{3
R}\right)\right]} \left<\varphi^2\right>_{IR}^{(\Phi_n(R)\simeq
0)},
\end{equation}
where\cite{ultimo}
\begin{equation}\label{asim6}
\left<\varphi^2\right>_{IR}^{(\Phi_n(R)\simeq 0)}=\frac{2^{2\nu
-3}}{3-2\nu} \frac{\Gamma^{2}(\nu)}{\pi^3}H^{2\nu
-1}e^{-(3-2\nu)Ht} k_{H}^{3-2\nu} =\frac{2^{2\nu-3} H^2
\Gamma^2(\nu)}{\pi^3 (3-2\nu)\left(12\xi
_{1}+\frac{9}{4}\right)^{\nu-3/2}}.
\end{equation}
Finally, we can write the power spectrum of
$\left<\varphi^2\right>_{SS}$, making $k_R = 2\pi/ R$ and $k_p =
2\pi / \lambda_p$
\begin{equation}
\left.{\cal P}(k_R)\right|_{\left<\varphi^2\right>_{SS}} \sim
e^{-\left[\frac{1}{2}\left(n \frac{k_R}{k_p}\right)^2
\left(1+\frac{16n}{3}\frac{k_R}{k_p}\right)\right]} k^{3-2\nu}_R.
\end{equation}
In the figure (\ref{f2}) we have plotted the coupling parameter
$\xi^{n,l}_2(R)$ in eq. (\ref{coup}) on astrophysical scales [$R_*
< R < 500 \  R_*$], for $n=10$, $l=0$ and $A=10^{12}$. Notice that
we have used $\varphi_{k_R}(R)$ given in eq. (\ref{3c}) and
$\Phi_n(R)$ given in eq. (\ref{6}). It is evident that the
coupling parameter becomes more important as $R$ increases (i.e.,
for bigger scales). In the figure (\ref{f3}) we shows the power
spectrum $\left.{\cal
P}(k_R)\right|_{\left<\varphi^2\right>_{SS}}$ as a function of the
wavenumber $k_R$ on astrophysical scales. There are three indices
that are important on the whole spectrum. Two of these dominate on
astrophysical scales. The first one for $k_R > 0.4$, the second
one on the range $0.01 < k_R < 0.4$. The third index $n_s \simeq
1$ dominates on cosmological scales ($k_R \ll 0.01$). We have used
$n_s=0.964$ (which corresponds to $\nu=1.518$) in all the
graphics.

\section{Final comments}

We have studied the power spectrum on astrophysical scales of the
$\varphi^2$-expectation value. The model here studied predicts a
spectrum that agrees qualitatively with experimental
data\cite{PDG}. An interesting result is that we detect two
different sectors with different power indices, which dominate on
different astrophysical scales. The third index is $n_s$, which is
relevant on cosmological scales. Another interesting result is the
periodicity of the coupling $\xi^{n,l}_2(R)$. Notice that its
amplitude increases with $R$. This coupling should be responsible
for the correlation of the galaxies on astrophysical scales.

For simplicity, in this letter we are considered a de Sitter
expansion, where the Hubble parameter $H$ is a constant. However,
the formalism could be extended to whatever $H=H(t)$.
The possibility of having a dynamical foliation 
using only 5D (in models without gravitational sources), was
explored\cite{nintro1}.
In our model the fifth dimension is 
responsible for the 4D de Sitter expansion, which is physically
driven by the inflaton field $\varphi$. From the physical point of
view, the sixth dimension is responsible for the spatial curvature
induced by the mass of the white hole (located at $R=0$). In more
general terms, the fifth dimension is physically related to the
vacuum energy density which is the source of the effective 4D
global inflationary expansion whereas the sixth one induces local
gravitational sources.

\vskip .2cm \centerline{\bf{Acknowledgements}} \vskip .2cm The
author
acknowledges CONICET and UNMdP (Argentina) for financial support.\\

\end{document}